\documentclass[pdflatex,sn-mathphys-num]{sn-jnl}

\usepackage{graphicx}%
\usepackage{multirow}%
\usepackage{amsmath,amssymb,amsfonts}%
\usepackage{amsthm}%
\usepackage{mathrsfs}%
\usepackage[title]{appendix}%
\usepackage{xcolor}%
\usepackage{textcomp}%
\usepackage{manyfoot}%
\usepackage{booktabs}%
\usepackage{algorithm}%
\usepackage{algorithmicx}%
\usepackage{algpseudocode}%
\usepackage{listings}%

\theoremstyle{thmstyleone}%
\theoremstyle{thmstyletwo}%

\theoremstyle{thmstylethree}%

\begin{document}

\title[Human-AI for 2D/3D Registration Quality]{Human-AI Collaboration and Explainability for 2D/3D Registration Quality Assurance}


\author*[1]{\fnm{Sue Min} \sur{Cho}}\email{scho72@jhu.edu}

\author[1]{\fnm{Alexander} \sur{Do}}

\author[1]{\fnm{Russell H.} \sur{Taylor}}

\author[1]{\fnm{Mathias} \sur{Unberath}}

\affil[1]{\orgname{Johns Hopkins University}, \city{Baltimore}, \state{MD}, \country{USA}}

\abstract{\textbf{Purpose:} As surgery increasingly integrates advanced imaging, algorithms, and robotics to automate complex tasks, human judgment of system correctness remains a vital safeguard for patient safety. A critical example is 2D/3D registration, where small registration misalignments can lead to surgical errors. Current visualization strategies alone are insufficient to reliably enable humans to detect these misalignments, highlighting the need for enhanced decision-support tools.

\textbf{Methods:} We propose the first artificial intelligence (AI) model tailored to 2D/3D registration quality assessment, augmented with explainable AI (XAI) mechanisms to clarify the model’s predictions. Using both objective measures (e.g., accuracy, sensitivity, precision, specificity) and subjective evaluations (e.g., workload, trust, and understanding), we systematically compare decision-making across four conditions: AI-only, Human-only, Human+AI, and Human+XAI.

\textbf{Results:} The AI-only condition achieved the highest accuracy, whereas collaborative paradigms (Human+AI and Human+XAI) improved sensitivity, precision, and specificity compared to standalone approaches. Participants experienced significantly lower workload in collaborative conditions relative to the Human-only condition. Moreover, participants reported a greater understanding of AI predictions in the Human+XAI condition than in Human+AI, although no significant differences were observed between the two collaborative paradigms in perceived trust or workload.

\textbf{Conclusion:} Human-AI collaboration can enhance 2D/3D registration quality assurance, with explainability mechanisms improving user understanding. Future work should refine XAI designs to optimize decision-making performance and efficiency. Extending both the algorithmic design and human–XAI collaboration elements holds promise for more robust quality assurance of 2D/3D registration.}




\maketitle

\section{Introduction}\label{sec1}
Surgery is undergoing a profound digital transformation, evolving from procedures performed solely by human experts to those guided by sophisticated imaging and advanced algorithms, assisted or automated through robotic technology. Yet, even as technology reshapes how modern surgery is delivered, human judgment remains indispensable for ensuring correct system function, and thus, patient safety~\cite{rivero2024autonomous}. The emerging concept of human-centered assurance underscores the need to integrate human operators into complex, technology-assisted workflows~\cite{cho2023visualization}. Still, the precise roles and responsibilities of these operators are far from clearly defined~\cite{fiorini2022concepts}. As surgical platforms become more automated, new tasks---such as verifying advanced algorithmic outputs---will increasingly fall to staff members or specialists who may not hold traditional clinical titles. Consequently, understanding how these ``operators'' perceive, interpret, and act on algorithmic information is essential for robust safety assurance.

This challenge is exemplified in the assurance of image-based navigation systems, where humans must verify the correctness of algorithmic alignments that directly affect surgical precision. In semi- and fully autonomous minimally invasive surgery, human oversight remains indispensable even as autonomy increases. Although machine intelligence promises to reduce errors and enhance precision, the responsibility for ensuring patient safety remains squarely with those in the operating room. Surgical robots perform both routine maneuvers, such as navigating through unoccupied space, and critical tasks, such as bone drilling or implant placement, where even millimeter-scale errors can cause permanent damage. These high-stakes ``branching decisions'' require not only accurate intraoperative guidance but also sufficient transparency for human operators to confidently validate algorithmic outputs.

These assurance challenges are vividly illustrated by 2D/3D registration, a key enabler in image-guided navigation that aligns intraoperative 2D fluoroscopic images with pre- or intraoperative 3D volumes. Indeed, image-based navigation---where the relative poses of surgical plans, instruments, and anatomy are estimated directly from image data---has long been heralded as the future of navigated surgery~\cite{mirota2011vision,markelj2012review}. However, despite significant gains in the accuracy and autonomy of 2D/3D registration through both optimization-based and deep learning approaches~\cite{unberath2021impact}, ensuring quality and safety remains a central challenge. Because image-based navigation is most frequently used in delicate anatomy, such as the spine, even small misalignments (on the order of several millimeters) can lead to critical deviations in tool placement or implant positioning.

To address this gap, this work integrates algorithmic, explainable, and human-centered approaches to enable more reliable and interpretable quality assurance in image-guided surgery. Specifically, we: 1) introduce the first learning-based model for 2D/3D registration quality assurance, trained to predict registration success, defined by a mean target registration error (mTRE) below 2mm, 2) integrate the model with explainability mechanisms---Grad-CAM for spatial attribution and conformal prediction for calibrated confidence---providing interpretable, case-specific rationales that support human oversight, and 3) conduct a systematic comparison across four decision-making conditions (AI-only, Human-only, Human+AI, Human+XAI), quantifying their respective strengths, limitations, and effects on task performance and user experience in safety-critical verification.

\section{Related Work}
\subsection{Ensuring Spatial Alignment in AR/MR for Surgery}
Accurate spatial alignment between virtual models and physical anatomy is essential for AR/MR-assisted surgery. Misalignments during this process can jeopardize both safety and precision. Researchers have explored perception-driven methods that leverage human interaction and visualization paradigms to detect and correct spatial errors. AR-guided ``registration sanity check'' protocols in head and face surgery enable operators to visually inspect virtual landmarks to validate alignment accuracy prior to intervention~\cite{condino2023registration}. Efforts to improve spatial misalignment detection have also focused on evaluating conventional MR visualization paradigms, which have proven less effective for detecting subtle alignment errors~\cite{gu2022impact}. Innovations in interactive methodologies, such as reflective-AR displays, have demonstrated improved precision by enabling surgeons to align virtual and physical environments more effectively~\cite{fotouhi2020reflective}.

While these AR/MR studies illustrate the importance of perceptual validation, they primarily address 3D-to-3D alignment scenarios. In contrast, 2D-to-3D registration, also critical in image-guided navigation, poses distinct geometric and perceptual challenges that remain underexplored. 

\subsection{Quality Assurance for 2D/3D Registration}
Human-based verification methods have been proposed to involve operators in the assessment of 2D/3D registration outcomes~\cite{cho2023visualization,cho2025feeling}. 
Studies using both existing and novel visualization paradigms showed that, on average, users can discern different levels of registration error; however, consistently and reliably detecting misalignments remains insufficient. Beyond human verification, non-learning-based probabilistic methods have sought to quantify registration uncertainty and correlate it with geometric accuracy~\cite{cho2025uncertainty}. Although this approach provides valuable confidence estimates, the relationship between uncertainty and true registration error is nonlinear and insufficient for error-proof validation. 

Together, these studies have established the foundations for quality assurance for 2D/3D registration, but indicate that it remains an open and difficult challenge, pointing to the need for a complementary approach that leverages the strengths of both humans and algorithms.

Our study further progresses this area by introducing a learning-based approach to registration quality assessment, designed not only to predict success but also to explain its reasoning and confidence. By comparing AI-only, Human-only, Human+AI, and Human+XAI conditions, this study explores the potential of human-AI collaboration to achieve safer and more consistent assurance of algorithmic outputs in technology-assisted surgery.

\begin{figure}[t]
\centering
\resizebox{0.8\textwidth}{!}{\includegraphics{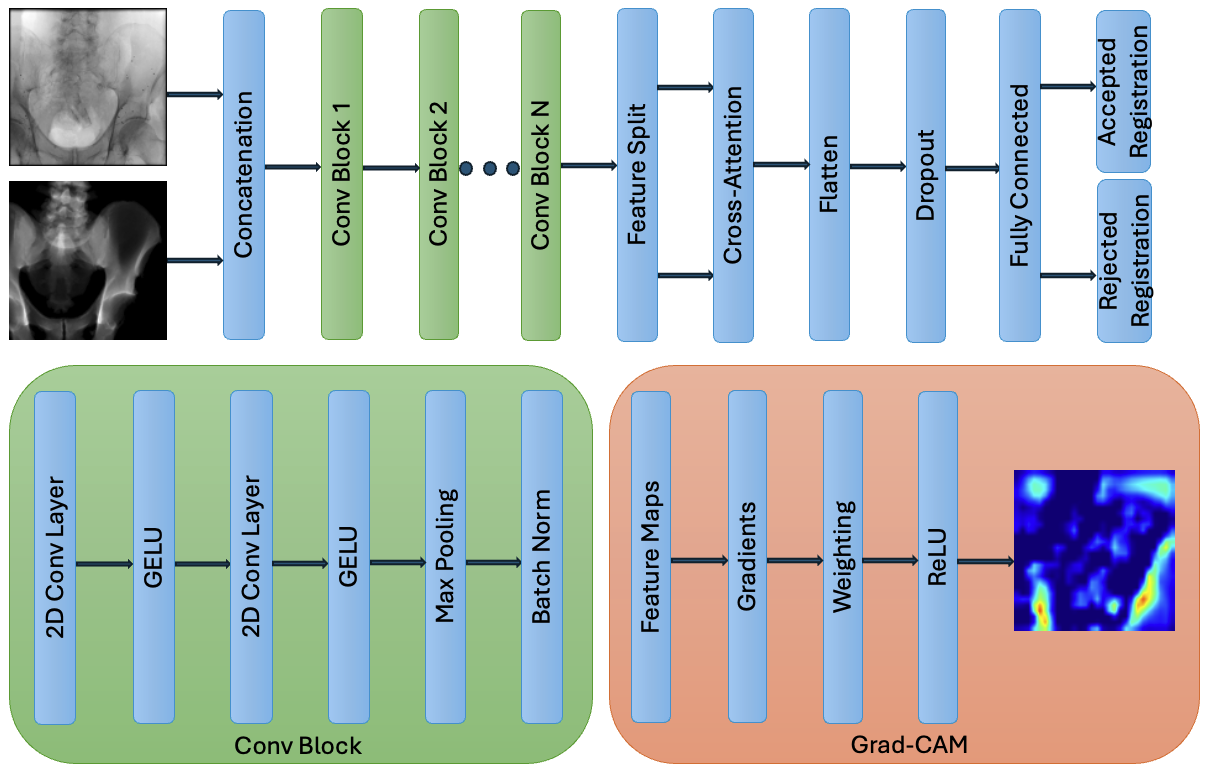}}
\caption{Overview of Proposed Model Architecture (The X-ray, DRR, and Grad-CAM output corresponds to Specimen ID: 18-2800, Projection ID: 0, Sample ID: 24)}
\label{fig:architecture}
\end{figure}

\section{Methods}
\subsection{Algorithmic Framework}\label{algorithm}
\subsubsection{Dataset and Preprocessing}\label{dataset}
Fluoroscopic images of the pelvis were obtained from a public dataset~\cite{T1/IFSXNV_2020}, which included five unique specimens. This dataset contained real fluoroscopic projections, CT scans from cadaveric specimens, and the corresponding ground truth poses for the real projections. Using DeepDRR~\cite{unberath2018deepdrr}, additional simulated projections were generated from the CT scans, resulting in a total of 200 projections (real + simulated) per specimen. Each image was paired with its reference ground truth poses and anatomical landmarks. We uniformly sampled 6 degrees of freedom (6DoF) for registration initialization and used an open-source 2D/3D single-view registration algorithm~\cite{grupp2018patch} to generate 100 registration results per projection. For each result, the offset and the corresponding digitally reconstructed radiograph (DRR) were saved, yielding a dataset of 100,000 DRRs and 1,000 fluoroscopic projections. To evaluate registration accuracy, we computed the mTRE based on the ground-truth 3D landmarks and the 3D points transformed by the offsets. Registrations were classified as successful if the mTRE was below 2mm; otherwise, they were considered failed. For partitioning the data into training, validation, and test sets, we employed leave-one-out cross-validation (LOOCV) at the specimen level. In each fold, three specimens were used for training, one for validation, and one for testing.

In order to deal with a class imbalance where there were far more failed registrations, non-geometric data augmentations were applied to increase the number of successful registrations during training. In particular, random additive Gaussian noise (with prob=0.6), Gaussian blur (with prob=0.4), and brightness/contrast (with prob=0.8) were applied. Projections with $90\%$ of registrations failed were removed. The ratio of successful to failed registrations was adjusted from approximately 1:2 to nearly 1:1, effectively reducing the class imbalance in the training set.

\subsubsection{Model Architecture and Training Procedure}\label{model}
An overview of the model architecture can be seen in Fig.~\ref{fig:architecture}. We utilize an early fusion approach where X-ray and DRR images are concatenated along the channel dimension, doubling the first convolution block's channel size. Each convolution block consists of two convolutional layers ($3\times 3$ kernels) with GELU activation followed by each layer, max pooling ($2\times 2$ kernels with stride $2$), and batch normalization. After feature extraction, the final feature map is split channel-wise into two views that interact through bidirectional cross-attention. The cross-attended features are then fused with an averaging operation, flattened, and processed through a dropout layer before a fully connected layer produces the binary prediction classes. 

The training procedure was optimized using Optuna~\cite{akiba2019optuna}, which determined the optimal hyperparameters for the model. The optimal hyperparameters found were:\texttt{learning\_rate=0.0002, weight\_decay=0.00005}, and \texttt{batch\_size=16}. Binary cross-entropy was used as the loss function during training, ensuring efficient learning across tasks. The model was implemented in Pytorch 2.3.1 and trained on an NVIDIA Titan RTX GPU.

\subsubsection{Explainability Mechanisms}\label{explainability}
To visualize the spatial locations of the X-ray images that are most important to the model's predictions, Grad-CAM~\cite{selvaraju2017grad} is used (Fig. \ref{fig:architecture}). Hooks are set at the last convolutional layer in the backbone to capture feature maps and gradients flowing back to this layer. During inference time, for each X-ray and DRR image input, we spatially average the gradients to obtain channel weights, compute the weighted sum of the feature maps, and apply ReLU to show only positive contributions.

Conformal prediction is used to obtain statistical insights about the model's prediction. We compute noncomformity scores on a calibration set measuring the difference between prediction and ground truth and choose the $1-\alpha$ quantile as the threshold. Prediction sets are constructed on test examples by including all possible labels (accepted or rejected registration) that yield nonconformity scores less than or equal to the threshold. A single label indicates the model's certainty whereas two labels indicates uncertainty.

\begin{figure}[t]
\centering
\resizebox{1.0\textwidth}{!}{\includegraphics{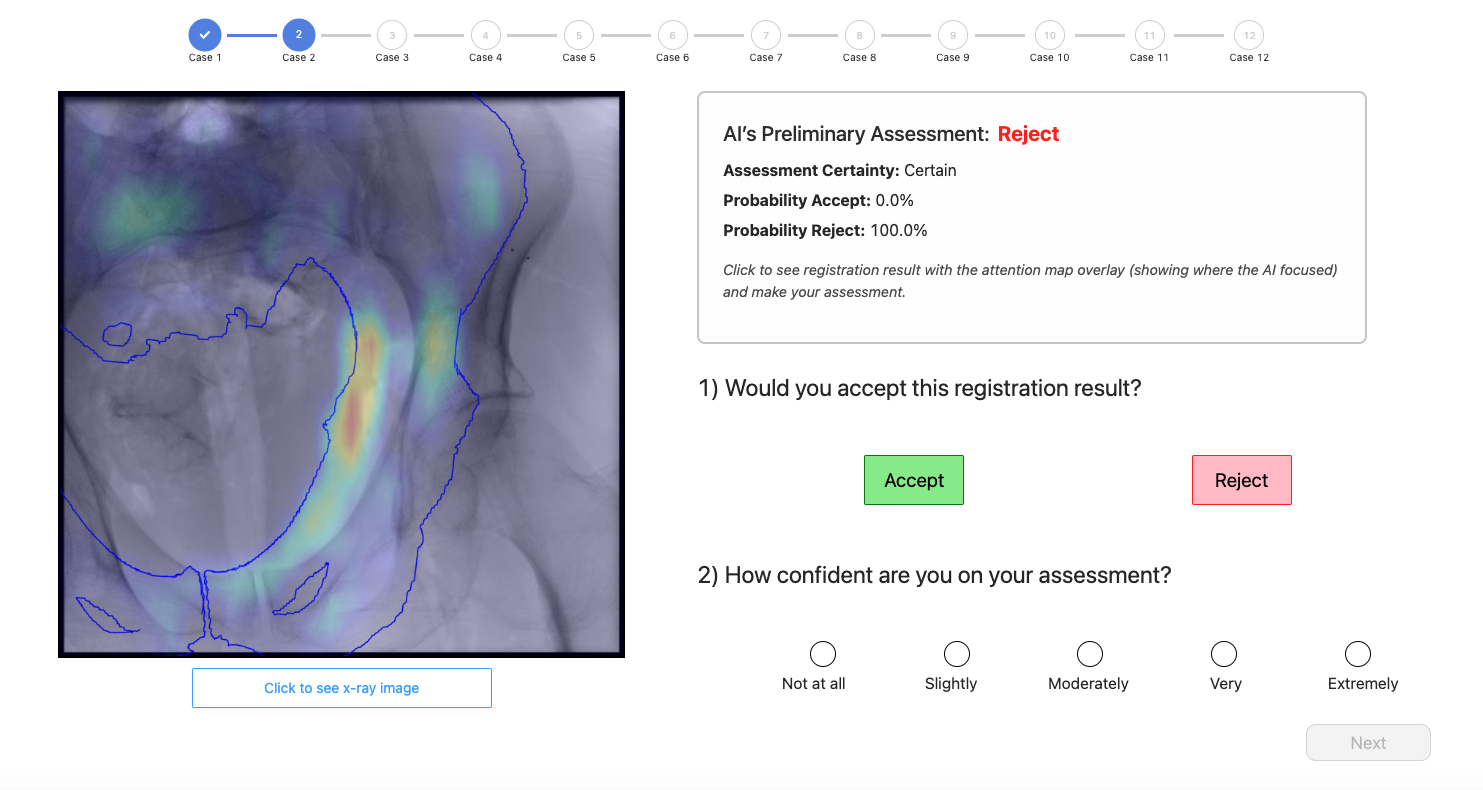}}
\caption{User interface in the Human–XAI condition. Participants viewed the AI’s classification decision (``Accept'' or ``Reject'') along with confidence scores and Grad-CAM heatmaps overlaid on the X-ray images and registration overlay, highlighting spatial regions that influenced the AI’s judgment.}
\label{fig:xai}
\end{figure}

\subsection{Human-AI User Study Design}\label{userstudy}
18 participants (8 males, 10 females), all with a STEM background, were recruited for the user study. This demographic was chosen to represent potential ``operators'' who may be involved in safety assurance and validation of computer-assisted intervention systems. The study design was approved by the local IRB, and informed consent was obtained from all participants prior to their involvement.

The user interface (Fig. \ref{fig:xai}) was implemented using a Next.js-based browser platform for data collection. At the start of the study, participants received instructions and were presented with example cases demonstrating various registration offsets. The study used a within-subjects design where participants completed three conditions: Human-Only (No AI), Human+AI (AI assistance without explanations), and Human+XAI (AI assistance with explainability). The Human-only condition was always presented first to ensure that participants established a baseline understanding of the task without AI influence. The order of the Human+AI and Human+XAI conditions was randomized across participants to counterbalance potential ordering effects. During each condition, participants completed 12 assessment tasks. For each task, an X-ray image and its registration overlay were shown, and participants were asked to assess the alignment as ``Accept'' or ``Reject,'' and provide their confidence levels using a 5-point Likert scale, from ``Not at all'' to ``Extremely.''

In the Human+AI condition, participants were additionally shown the AI's classification decision (Accept or Reject). In the Human+XAI condition, participants were shown the AI's confidence scores for its decision and Grad-CAM heatmaps, which were overlayed on the X-ray images to highlight spatial regions influencing the AI's predictions (Fig. \ref{fig:xai}). The order of X-ray images and registration offsets was randomized for each participant using an automated randomization procedure integrated into the experimental platform. After completing each condition, participants completed a short survey evaluating perceived workload and their experience with the AI assistance, if given. A post-study survey gathered demographic information.

\subsection{Evaluation}

\subsubsection{Performance Metrics}\label{performance_metrics}
Performance was quantitatively evaluated using standard classification metrics: \textit{accuracy}, \textit{sensitivity} (true positive rate, TPR), \textit{precision}, and \textit{specificity} (true negative rate, TNR). Participants judged registration quality by choosing either ``Accept'' or ``Reject,'' which were compared against the ground truth. A correct ``Accept'' corresponds to a \textit{true positive} (TP) and a correct ``Reject'' to a \textit{true negative} (TN), while incorrect decisions constitute \textit{false positives} (FP) and \textit{false negatives} (FN). 

\textit{Accuracy} reflects the overall fraction of correctly classified registrations (both accepted and rejected). \textit{Sensitivity} (TPR), the fraction of correct registrations successfully accepted, captures the ability to identify successful registrations without incorrectly rejecting them. \textit{Precision}, the fraction of accepted registrations that are actually correct, reflects decision quality when the system or operator chooses to accept, that is, how reliably accepted cases are truly correct. In contrast, \textit{specificity} (TNR) measures the fraction of incorrect registrations that are correctly rejected, capturing the ability to identify and dismiss failed registrations and thus prevent false acceptances in safety-critical settings.

To systematically evaluate the Human+AI and Human+XAI conditions based on AI-derived outcomes, we constructed a balanced subset containing equal numbers of TP, TN, FP, and FN cases. While AI models can be evaluated on large datasets, user studies are limited by participant capacity; the balanced subset ensures a representative yet manageable set of cases for human evaluation. Although the balanced subset clarifies user responses to various AI errors, it does not reflect the actual error distribution. Thus, we computed approximate ``real-world'' metrics by weighting each category’s fraction-correct by its prevalence in the entire test set (\textit{TP}=22.6\%, \textit{TN}=53.4\%, \textit{FP}=18.8\%, \textit{FN}=5.1\%).

\subsubsection{Subjective Measures}
Subjective evaluations were collected to assess participants’ perceived workload and their experience with AI assistance. Perceived workload was measured using the NASA-TLX questionnaire~\cite{hart1988development}. Additionally, participants rated trust in, usefulness of, and understanding of AI assistance in both Human+AI and Human+XAI conditions using a Likert-scale questionnaire (1: strongly disagree to 5: strongly agree). These subjective evaluations complement quantitative performance metrics by offering insights into participants’ experiences; details of the survey items can be found in the supplementary material.

To analyze differences across conditions, pairwise comparisons were conducted within subjects. Normality of paired differences was assessed with the Shapiro-Wilk test. For normally distributed differences ($p > 0.05$), paired t-tests were used; otherwise, Wilcoxon signed-rank tests were applied. Statistical analyses were performed using Python’s scipy library (v1.13.1).

\begin{table}[t]
\centering
\caption{Descriptive Statistics (Mean $\pm$ SD) for Performance Metrics Across Conditions}
\label{tab:weighted_metrics_summary}
\begin{tabular}{lcccc}
\hline
\textbf{Condition} & \textbf{Accuracy}     & \textbf{Sensitivity (TPR)}    & \textbf{Precision} & \textbf{Specificity (TNR)}       \\
\hline
AI-Only & $\mathbf{0.757 \pm 0.033}$    & $0.743 \pm 0.318$    & $0.575 \pm 0.098$     & $0.740 \pm 0.182$      \\
Human-Only & $0.547 \pm 0.158$    & $0.736 \pm 0.285$   & $0.531 \pm 0.248$      & $0.701 \pm 0.199$      \\
Human+AI & $0.677 \pm 0.125$   & $\mathbf{0.964 \pm 0.061}$   & $0.659 \pm 0.185$      & $0.848 \pm 0.079$      \\
Human+XAI & $0.678 \pm 0.124$  & $0.877 \pm 0.239$    & $\mathbf{0.675 \pm 0.294}$     & $\mathbf{0.865 \pm 0.099}$      \\
\hline
\end{tabular}
\end{table}

\section{Results}
\subsection{Performance Across Decision-Making Conditions}
AI-only was evaluated using cross-validation, with detailed evaluation metrics across the five folds provided in the supplementary material. Human-involved conditions were analyzed using the weighted metrics as described in~\ref{performance_metrics}. Table~\ref{tab:weighted_metrics_summary} summarizes the performance metrics for each condition.

The AI-only condition achieved the highest \textit{accuracy} ($0.757 \pm 0.033$). Among the human-inclusive conditions, Human+AI ($0.677 \pm 0.125$) and Human+XAI ($0.678 \pm 0.124$) achieved higher accuracies than Human-only ($0.547 \pm 0.158$). Accuracy in Human+XAI was slightly higher than Human+AI, though the difference was minimal. For \textit{sensitivity} (TPR), which measures the fraction of successful registrations correctly accepted,  Human+AI achieved the highest value ($0.964 \pm 0.061$), followed by Human+XAI ($0.877 \pm 0.239$). The AI-only and Human-only conditions showed lower sensitivities ($0.743 \pm 0.318$ and $0.736 \pm 0.285$, respectively). In terms of \textit{precision}, which reflects the fraction of accepted registrations that are truly correct (decision quality for accepted cases), Human+XAI achieved the highest mean value ($0.675 \pm 0.294$), followed closely by Human+AI ($0.659 \pm 0.185$). The AI-only and Human-only conditions showed lower precision ($0.575 \pm 0.098$ and $0.531 \pm 0.248$, respectively). For \textit{specificity} (TNR), which measures the fraction of failed registrations that are correctly rejected (key for avoiding false acceptances in safety-critical systems),  Human+XAI achieved the highest mean ($0.865 \pm 0.099$), followed by Human+AI ($0.848 \pm 0.079$), AI-only ($0.740 \pm 0.182$), and Human-only ($0.701 \pm 0.199$).

\begin{figure}[t]
\centering
\resizebox{0.95\textwidth}{!}{\includegraphics{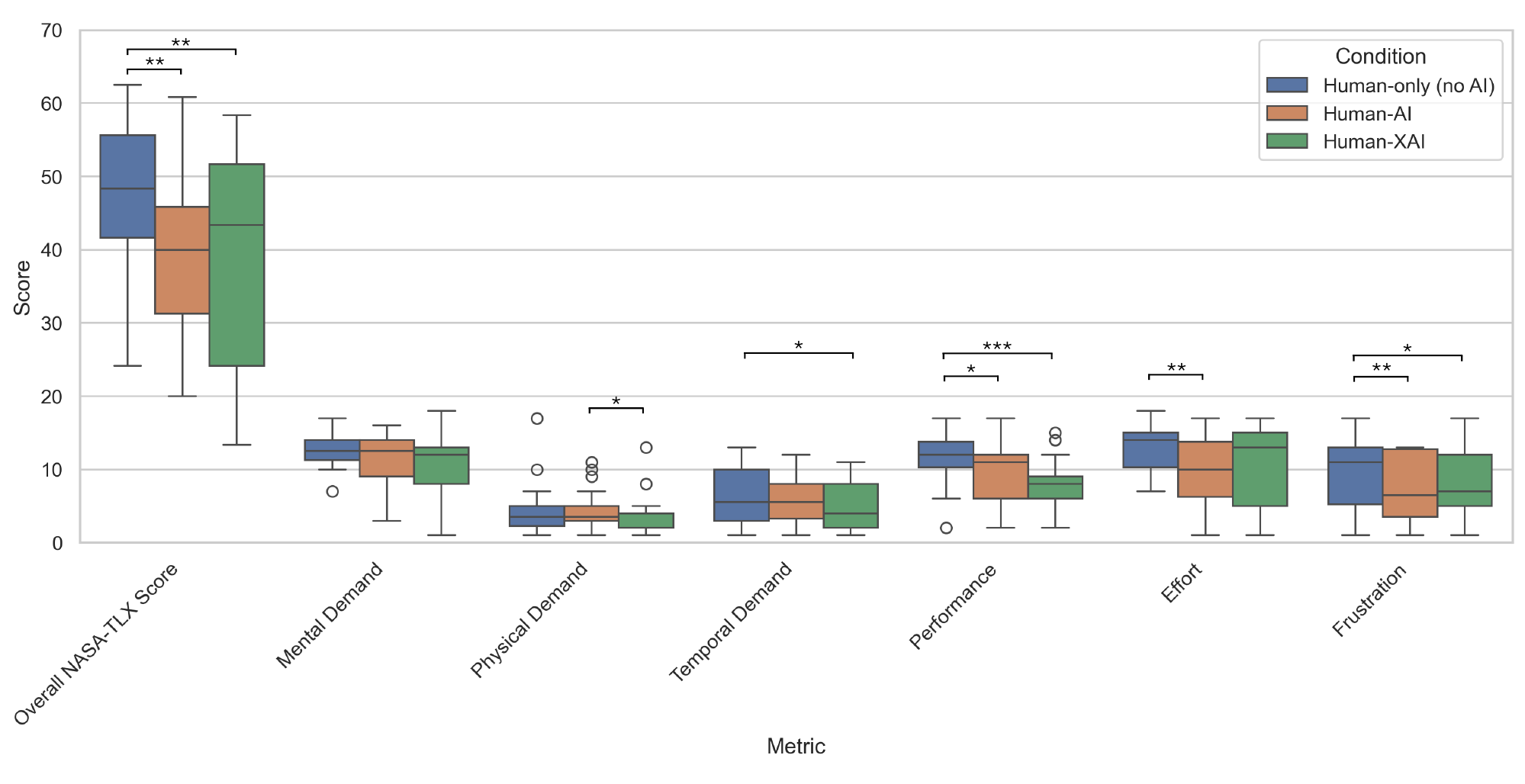}}
\caption{Box plots of NASA-TLX scores across three conditions: (1)Human-only, (2)Human+AI, and (3)Human+XAI. Lower scores indicate lower perceived workload. *$p<0.05$, **$p<0.01$, *** $p<0.001$}
\label{fig:NASA}
\end{figure}

\subsection{Perceived Workload Across Human-AI Conditions}
Figure~\ref{fig:NASA} presents NASA-TLX scores measured across the three experimental conditions, with pairwise statistical results summarized in the supplementary material. Overall perceived workload, measured by the aggregate NASA-TLX score, decreased substantially in both AI-assisted conditions compared to Human-only. Specifically, workload decreased significantly in both Human+AI (\(p = 0.0046\)) and Human+XAI (\(p = 0.0055\)) when compared to Human-only, and no significant difference was observed between Human+AI and Human+XAI (\(p = 0.3596\)). Assessment of specific workload dimensions using pairwise testing yielded further insights. Temporal demand was significantly reduced in the Human+XAI condition compared to Human-only (\(p = 0.0407\)), and Effort similarly decreased in the Human+AI condition compared to Human-only (\(p = 0.0060\)). Frustration levels decreased significantly in both Human+AI (\(p = 0.0067\)) and Human+XAI (\(p = 0.0325\)) conditions compared to Human-only. Performance assessments, measured by participants' self-perceived effectiveness in completing the task, were significantly higher in both Human+AI (\(p = 0.0208\)) and Human+XAI (\(p = 0.0005\)) conditions relative to Human-only. No statistically significant differences were observed between Human+AI and Human+XAI conditions in Performance metrics (\(p = 0.0938\)). Mental Demand showed no significant variances between any of the conditions. Physical Demand was marginally lower for Human+XAI (\(p = 0.0492\)) compared to Human+AI.

\subsection{Perceived Qualities of AI Assistance}
Participants’ subjective evaluations of AI assistance in Human+AI and Human+XAI conditions are summarized in the supplementary material. No significant differences were identified for the metrics assessing perceived trust (\(p = 0.276\)), impact (\(p = 0.707\)), or helpfulness (\(p = 0.472\)). However, perceived understanding was significantly higher in the Human+XAI condition (\(p = 0.030\)), indicating that explainability mechanisms enhanced participants’ comprehension of AI predictions.

\section{Discussion and Conclusion}
In this work, we presented the first learning-based model for assessing 2D/3D registration quality, integrated it with explainability mechanisms, and compared four decision-making paradigms (AI-only, Human-only, Human+AI, Human+XAI).

The performance evaluation used standard classification metrics---accuracy, sensitivity (TPR), precision, and specificity (TNR)---described in Section~\ref{performance_metrics}. The AI-only condition achieved the highest overall accuracy. However, the collaborative paradigms (Human+AI and Human+XAI) outperformed standalone conditions in sensitivity (TPR), precision, and specificity (TNR), demonstrating the added value of integrating human judgment with AI support. Human+AI and Human+XAI showed markedly higher sensitivity, indicating improved detection of successful registrations. Human+XAI achieved the highest precision and specificity, suggesting that AI guidance and explainability effectively supported participants in avoiding false acceptance of failed registrations and correctly rejecting them. Clinically, these gains are meaningful, as false acceptances of misaligned registrations pose greater safety risks than false rejections of correct alignments. While Human+XAI provided modest performance advantages over Human+AI, the differences were small, indicating room for improvement in current explainability mechanisms.

Regarding subjective measures, participants reported significant reductions in perceived workload in both Human+AI and Human+XAI conditions compared to Human-only, suggesting that AI support effectively reduces cognitive effort in decision-making tasks. However, no significant difference in workload was identified between Human+AI and Human+XAI, indicating that explainability mechanisms did not add measurable cognitive burden. While participants in the Human+XAI condition provided higher ratings for understanding, no significant improvements over Human+AI were observed for perceived trust, impact, or helpfulness of AI assistance. These subjective evaluation results may have been influenced by the use of a balanced subset of AI-derived outputs. By ensuring equal numbers of correct and incorrect AI predictions in the balanced set, participants’ perception of AI effectiveness may not have fully reflected real-world AI performance, where the distribution of errors may differ. While this design allowed a controlled evaluation of human responses to correct and incorrect predictions, it may have impacted participants' trust in the system overall.

In addition, we used a 2mm threshold to define successful registration. However, it is important to note that the clinically acceptable margin can be highly application-specific, for example, in spine procedures where required accuracy ranges from 0.0mm to 3.8mm~\cite{rampersaud2001accuracy}. Thus, performance metrics -- and ultimately, conclusions -- may shift based on clinical thresholds chosen for different anatomy or tasks, underscoring the need for adaptable standards in evaluating surgical registration systems.

Several avenues remain open for future exploration. First, the interaction between human operators and XAI was relatively static in our study. Future work could incorporate iterative or conversational interfaces, where operators can query the AI on uncertain areas for real-time feedback. Second, adopting gaze tracking~\cite{cho2024misjudging} and other physiological or behavioral signals could offer deeper insights into how operators detect misalignments or interpret AI explanations. Third, a wider array of algorithmic solutions should be explored to ensure robust performance in diverse clinical conditions

Overall, our findings highlight the dual need for accurate models and well-designed explanations and interactions that truly support human judgment in safety-critical contexts. As surgical technologies continue to advance, it remains essential to pursue a human-centered perspective, ensuring they augment rather than supplant the expertise and accountability of the operating room team.

\backmatter

\section*{Statements and Declarations}
\noindent \textbf{Conflict of Interest} All authors have no conflict of interest to declare.

\noindent \textbf{Ethical Approval} The study was approved by the Homewood Institutional Review Board (HIRB00013292) and conducted according to ethical standards.

\noindent \textbf{Informed Consent} Informed consents were obtained from all participants.

\noindent \textbf{Code and Data Availability} Code and data for our algorithmic framework are available at \href{https://github.com/jasminecho1008/XAI-2D3D-RegQuality}{https://github.com/jasminecho1008/XAI-2D3D-RegQuality}.

\bibliography{sn-bibliography}

\end{document}